\documentclass[epjCONF]{svjour}
\usepackage{graphicx}
\usepackage[small,sf]{caption}
\usepackage[varg]{txfonts} % Times fonts
\usepackage[latin1]{inputenc}
\session-title{%
19$^{\textnormal{\footnotesize th}}$ International %
IUPAP Conference on Few-Body Problems in Physics%
}
\begin{document}
\title{%
%Insert your title here:
Measurements of scattering observables for the $pd$ break-up reaction
}%
\author{%
% add email of the responsible author:
M. Eslami-Kalantari\inst{1,2}\fnmsep\thanks{\email{meslami@yazduni.ac.ir}}
\and %
H.R. Amir-Ahmadi\inst{1}
\and %
A. Biegun\inst{1}
\and %
I. Ga\v{s}paric\inst{3}
\and %
L. Joulaeizadeh\inst{1}
\and %
N. Kalantar-Nayestanaki\inst{1}
\and %
St. Kistryn\inst{4}
\and %
A. Kozela\inst{5}
\and %
H. Mardanpour\inst{1}
\and %
J.G. Messchendorp\inst{1}
\and %
H. Moeini\inst{1}
\and %
A. Ramazani-Moghaddam-Arani\inst{1,6}
\and %
S.V. Shende\inst{1}
\and %
E. Stephan\inst{7}
\and %
R. Sworst\inst{4}
}
\institute{%
KVI, University of Groningen, Groningen, The Netherlands
\and %
Faculty of Physics, Yazd University, Yazd, Iran
\and %
Rudjer Bo\v{s}kovi\'c Institute, Zagreb, Coratia
\and %
Institute of Physics, Jagiellonian University, Kracow, Poland
\and %
Henryk Niewodnicza\'nski, Institute of Nuclear Physics, Kracow, Poland
\and %
Department of Physics, Faculty of Science, University of Kashan, Kashan, Iran
\and %
Institute of Physics, University of Silesia, Katowice, Poland
}
\abstract{
High-precision measurements of the scattering observables such as cross sections and analyzing powers for the proton-deuteron elastic and break-up reactions have been performed at KVI in the last two decades and elsewhere to investigate various aspects of the three-nucleon force (3NF) effects simultaneously. In 2006 an experiment was performed to study these effects in $\vec{p}+d$ break-up reaction at 135 MeV with the detection system, Big Instrument for Nuclear polarization Analysis, BINA. BINA covers almost the entire kinematical phase space of the break-up reaction. The results are interpreted with the help of state-of-the-art Faddeev calculations and are partly presented in this contribution.
} %end of abstract
\maketitle
%
%
%----- Beginning of MAIN TEXT  ---------------------------------------
%

In the last few decades, several nucleon-nucleon potentials (NNPs) have been studied extensively to describe the properties of bound nuclear systems by comparing high-precision two-nucleon scattering data with modern potentials based on the exchange of bosons~\cite{Stoks1,Stoks2,Wiringa,Machleidt}. The modern NNPs reproduce the world database with a reduced $\chi^{2}$ close to one and have, therefore, been accepted as high-quality potentials. Deficiencies of theoretical predictions based on pair-wise nucleon-nucleon potentials have been observed in three-nucleon scattering observables. For example, exact solutions of the
Lippmann-Schwinger equations (Faddeev calculations)~\cite{glockle} solely based on modern NN interactions fail to describe high-precision differential cross sections of proton-deuteron elastic scattering at intermediate
energies obtained at many laboratories including
KVI~\cite{Ermisch1,Ermisch2,Ermisch3,Hamid}, Research Center for Nuclear Physics (RIKEN) \cite{Sakai,Kimiko1} and RCNP~\cite{Kimiko2}.
Calculations based on NNPs including $2\pi$-exchange type three-nucleon forces~(3NFs) remove this discrepancy for a large part and lead to a good description of the measured cross sections for energies below 100 MeV/nucleon. However, the description of spin observables such as vector and in particular the tensor analyzing powers is not satisfactory and an inclusion of 3NFs is not sufficient to remedy the observed discrepancies~\cite{Bieber,Ermisch1,Ermisch2,Ermisch3}.

The break-up reaction has a rich phase space which allows a systematic study of the 3NF. Predictions show that large 3NF effects can be expected at specific kinematical regions in the break-up reaction. For other phase space regions, the effect can be small, which make them, therefore, suitable for benchmark studies. In elastic and break-up reactions, precision data for a large energy interval for the differential cross section and analyzing power have come from recent experimental studies at KVI~\cite{Ermisch1,Ermisch2,Ermisch3,Hossein,Hamid,Ahmad,Mohammad}. Break-up cross sections and analyzing powers for a beam energy of $E^{d}_{lab}$ = 130 MeV have been published in Refs.~\cite{Biegun,Kistryn03,Kistryn05,Kistryn06,Stephan}. For a further study of 3NF effects, we have measured the break-up cross sections and vector analyzing powers for a proton beam energy of $E^{p}_{lab}$ = 135 MeV. In this experiment, cross sections and vector analyzing powers in $\vec{p}+d\rightarrow p+p+n$ reaction were measured using a polarized beam on a liquid-deuterium target~\cite{Nasser}. The results are compared with predictions derived from state-of-the-art Faddeev calculations and are partly reported here.

\begin{figure}[!h]
\begin{center}
\includegraphics[angle=0,width=.5\textwidth] {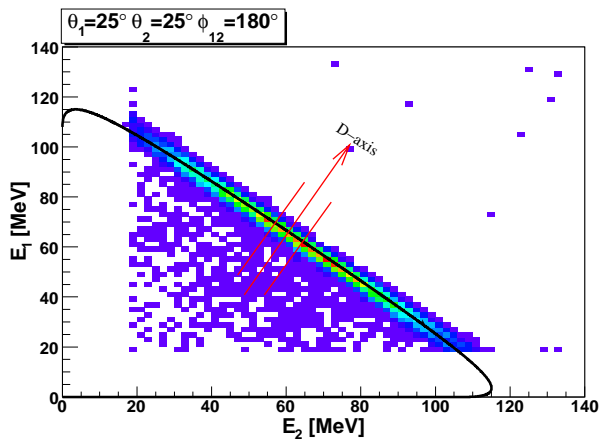}
\caption{The coincidence spectrum between the energies $E_{1}$ versus $E_{2}$ of the two protons registered at ($\theta_{1}$,$\theta_{2}$,$\phi_{12}$)=(25$^\circ$$\pm$2$^\circ$,25$^\circ$$\pm$2$^\circ$,180$^\circ$$\pm$5$^\circ$). The solid line shows the kinematical curve, the so-called $S$-curve calculated for the central values of the experimental angular ranges.}
\label{Eslami_fig_1}
\end{center}
\end{figure}

\begin{figure}[!h]
\begin{center}
\includegraphics[angle=0,width=.5\textwidth] {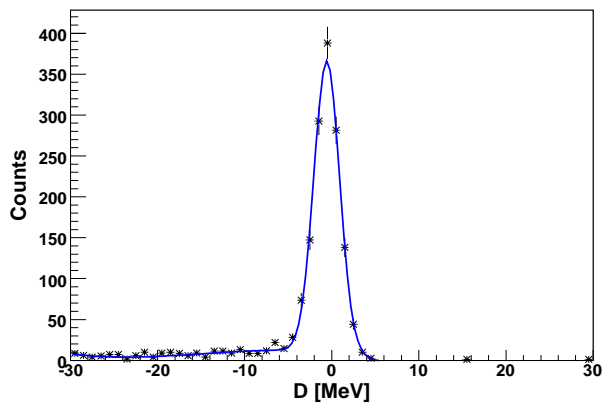}
\caption{The projection of the slice chosen in Fig.~\ref{Eslami_fig_1} along the D-axis. The solid line is the sum of a third-order polynomial background function and a Gaussian distribution representing the peak.}
\label{Eslami_fig_2}
\end{center}
\end{figure}

The proton-deuteron break-up experiment described in this paper was performed using the recently-developed detection system, BINA. The energy correlation between the two outgoing protons, $E_{1}$ vs.~$E_{2}$, as shown in Fig.~\ref{Eslami_fig_1}, was studied for a large number of kinematic configurations. To obtain the cross section and the analyzing power, several slices along the $S$-curve were made. The angular ranges for the integration of events was chosen to be $\Delta\theta_{1}$=$\Delta\theta_{2}$=$4^\circ$ and $\Delta\phi_{12}$=$10^\circ$, which were wide enough to have sufficient statistics, while variations of the cross section within these ranges are small. In this way, the experimental cross sections can be directly compared with the theoretical predictions calculated for the central values of a specific configuration.
The projection of the indicated region in Fig.~\ref{Eslami_fig_1} on the line perpendicular to the $S$-curve is shown in Fig.~\ref{Eslami_fig_2}. This spectrum contains break-up events around channel 0 and background from accidental events and contributions resulting from the hadronic interaction of particles in the scintillator material of the detector. This projection axis is denoted the D-axis. The crossing point of the $S$-curve with the D-axis defines the zero point.

The accidental background is located at channels higher than the main peak in Fig.~\ref{Eslami_fig_2} and is already subtracted exploiting the relative time-of-flight information between the two protons.
The contributions from hadronic interaction are located at lower channels on the D-axis which correspond to events below the $S$-curve in Fig.~\ref{Eslami_fig_1}. These events are mostly ``true'' break-up events. At this stage, this contribution is fitted using a third-order polynomial together with the main peak represented by a Gaussian function. The number of events underneath the Gaussian peak will be referred to as the number of break-up events and will be used to obtain the cross section and analyzing power after making efficiency corrections such as the one from hadronic interaction.

\begin{figure}[!h]
\centering
\includegraphics[angle=0,width=.5\textwidth]{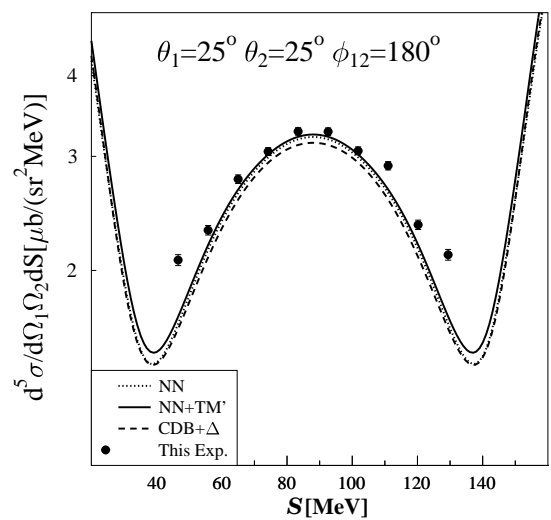}
\caption{The cross section is plotted as a function of $S$~[MeV] for the kinematical configuration, $(\theta_{1},\theta_{2},\phi_{12})$ =  $(25^{\circ},25^{\circ},180^{\circ})$. Lines represent Faddeev calculations from the Hannover-Lisbon, and Bochum-Krakow groups. The dotted line represents the cross section using the CD-Bonn two-nucleon potential, the solid line shows the CD-Bonn+TM' calculation. The dashed line represents the results of a calculation by the Hannover-Lisbon group, which is based on the extended CD-Bonn potential, including a virtual $\Delta$ excitation in a coupled-channel approach.}
\label{Eslami_fig_3}
\end{figure}

\noindent Figure~\ref{Eslami_fig_3} shows the cross section, $\frac{d^{5}\sigma}{d\Omega_{1}\, d\Omega_{2}\, dS}$[$\mu$b/(sr$^{2}$MeV)], as a function of $S$~[MeV] for the symmetric and coplanar configuration with $(\theta_{1},\theta_{2},\phi_{12})$ = $(25^{\circ},25^{\circ},180^{\circ})$.
In this figure, the lines represent the predictions by the Bochum-Krakow and Hannover-Lisbon theory groups \cite{witala,witala98,witala01,deltacolumb1,deltacolumb2}. The dotted line is the result of calculations using the CD-Bonn two-nucleon potential and the solid line presents the calculation including the three-body force, TM', as well. The dashed line represents the result of a calculation by the Hannover-Lisbon group, which is based on the extended CD-Bonn potential including a virtual $\Delta$ excitation in a coupled-channel approach.

In the following, we describe the determination of the vector analyzing power, $A_{y}$, by using a polarized proton beam and by measuring the induced asymmetry in the cross section. The relation between d$\sigma^{s}$= d$\sigma^{\uparrow,\downarrow}$ and the unpolarzied cross section, d$\sigma^{0}$, is
\begin{equation}
d\sigma^{\uparrow,\downarrow}=d\sigma^{0}(1+p^{\uparrow,\downarrow}\cdot A_{y}\cdot \cos\phi),
\label{Anaequation3}
\end{equation}
for an incoming proton beam with spin up ($\uparrow$) or spin down ($\downarrow$) and a vector analyzing power, $A_y$. Here $\phi$ is the angle between quantization axis for the beam polarization and the normal to the scattering plane in the laboratory frame of reference.
From the two cross sections, d$\sigma^{\uparrow}$ and d$\sigma^{\downarrow}$, with polarizations, $p^{\uparrow}$ and $p^{\downarrow}$, the analyzing power can be obtained from the asymmetry,

\begin{equation}
A_{y}\cos\phi=\frac{d\sigma^{\uparrow}-d\sigma^{\downarrow}}
{d\sigma^{\downarrow}p^{\uparrow}-d\sigma^{\uparrow}p^{\downarrow}}.
\label{Anaequation4}
\end{equation}

\begin{figure}[!h]
\centering
\includegraphics[angle=0,width=.5\textwidth]{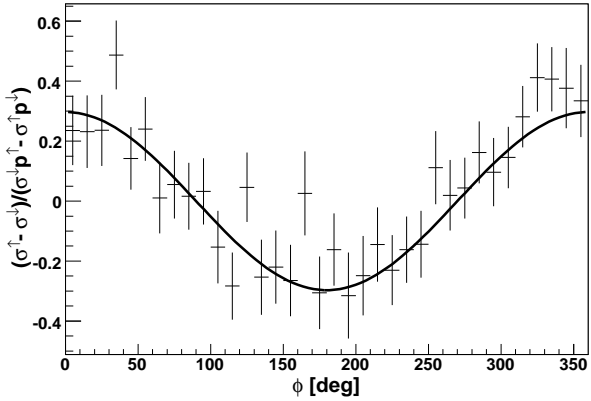}
\caption{The asymmetry, $A_{y}\cos\phi$, in the break-up reaction as a function of the azimuthal scattering angle of one of the protons, $\phi$.}
\label{Eslami_fig_4}
\end{figure}

\begin{figure}[!h]
\centering
\includegraphics[angle=0,width=.5\textwidth]{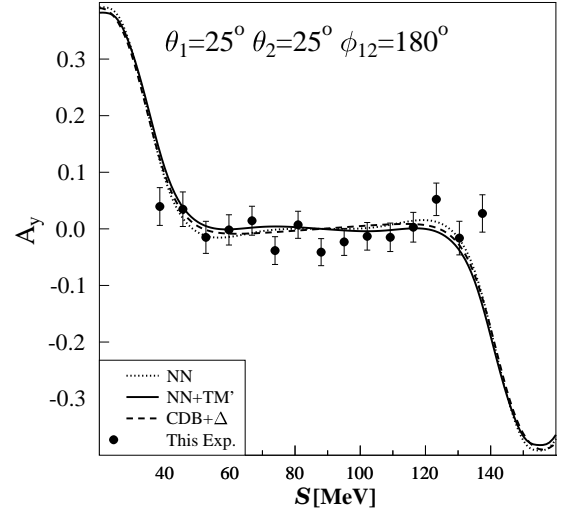}
\caption{Analyzing powers of the break up reaction for the coplanar kinematics, $(\theta_{1},\theta_{2},\phi_{12})$ = $(25^{\circ},25^{\circ},180^{\circ})$, as a function of $S$. For the explanation of the curves, see Fig.~\ref{Eslami_fig_3}.}
\label{Eslami_fig_5}
\end{figure}

The reaction asymmetry can be measured by exploiting the distribution of events obtained with beam polarizations up and down together with the values of the beam polarization in these two modes. This gives a periodic function in $\phi$ with an amplitude that corresponds to $A_{y}$. Figure~\ref{Eslami_fig_4} shows the asymmetry as a function of $\phi$ for a particular bin in $S$.
By exploiting the asymmetry distribution for each $S$-bin, the vector-analyzing power, $A_{y}$, is obtained for every kinematical configuration, $(\theta_{1},\theta_{2},\phi_{12})$. Figure~\ref{Eslami_fig_5} represents the analyzing-power for the configuration $(\theta_{1},\theta_{2},\phi_{12})$ = $($25$^{\circ}$,25$^{\circ}$,180$^{\circ})$. The various lines represent calculations from the Hannover-Lisbon and Bochum-Krakow theory gr\-oups as explained before.

\begin{figure*}[!htb]
\centering
\includegraphics[angle=0,width=.9\textwidth]{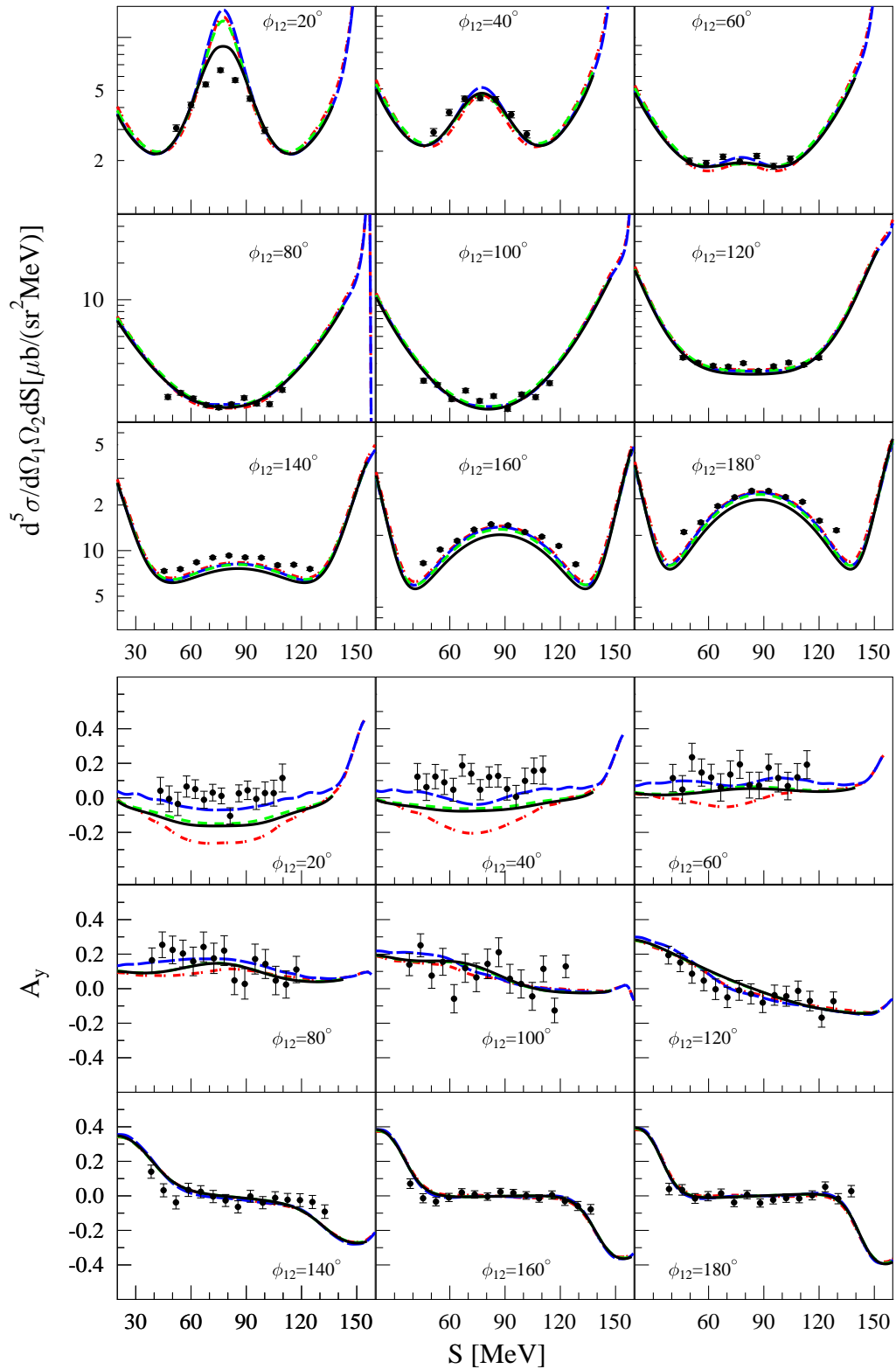}
\caption{The cross sections and the analyzing powers at
$(\theta_{1},\theta_{2})$ = $(25^{\circ}, 25^{\circ})$ as a function
of $S$ for different azimuthal opening angles. The error bars reflect only statistical uncertainties. The blue (long-dashed), red (dash-dotted), green (dashed), and black (solid) lines show predictions of Faddeev calculations using CDB
(NN), CDB+TM' (3NF), from the Bochum-Krakow group~\cite{witala,witala98,witala01}, CDB+$\Delta$ (3NF) and CDB+$\Delta$+Coulomb calculations from the Hannover-Lisbon group~\cite{deltacolumb1,deltacolumb2}, respectively.}
\label{Eslami_fig_6}
\end{figure*}

We determined the cross sections and analyzing powers for configurations in
which $14^{\circ}<\theta_{1,2}<30^{\circ}$, and the azimuthal opening angle,
$\phi_{12}$, varied from $20^{\circ}$ to $180^{\circ}$ in steps of
$20^{\circ}$. Fig.~\ref{Eslami_fig_6} shows the results of the fixed combination $(\theta_{1},\theta_{2})$ = ($25^{\circ},25^{\circ}$) with different azimuthal opening angles, $\phi_{12}$, as a function of $S$. The top panels depict the cross sections and the bottom panels show the analyzing powers.

The presented error bars in all the figures are statistical. The main
source of systematic
uncertainties for the break-up cross sections are: the uncertainty
in the target thickness (3.85$\pm$0.2 mm$\rightarrow\sim$5\%), the
correction for the hadronic reaction efficiency for both protons obtained via a
GEANT-3 simulation (92$\pm$3\%$\rightarrow$ $\sim$6\% for two
protons), the correction for the efficiency of the MWPC (92$\pm$1\%
$\rightarrow$ 2\% for two protons), and the correction for the
geometrical inefficiencies obtained via GEANT-3 simulations which is at most
 12\% for small azimuthal opening angles,
$\phi_{12}=20^{\circ}$, and at most 2\% for the larger azimuthal
opening angles. Altogether, by adding the systematic uncertainties in
quadrature, the maximum systematic uncertainty for cross sections at small
azimuthal opening angles is less than 14\% and at larger azimuthal opening
angles less than 9\%. The systematic error for the analyzing power
stems primarily from the uncertainty in the measurement of the beam
polarization via the proton-deuteron elastic-scattering reaction. For instance, the beam
polarization for the down-mode has been measured at a value of
$\sim$0.70 $\pm$ 0.04, which gives rise to a 6\% systematic uncertainty
in the analyzing power.

The predictions of the Faddeev calculations using different NN and 3NF
models are added to every panel with different line colors and styles. The blue (long-dashed),
red (dash-dotted), green (dashed), and black (solid) lines correspond to calculations based on CDB (NN),
CDB+TM' (3NF) from the Bochum-Krakow group~\cite{witala,witala98,witala01},
CDB+$\Delta$ (3NF),
and CDB+$\Delta$+Coulomb calculations from the Hann\-over-Lisbon
group~\cite{deltacolumb1,deltacolumb2}, respectively. Here, all theory curves
have been calculated in a fully non-relativistic framework with
non-relativistic observables and, therefore, the length of $S$ for
these calculations is slightly shorter than that in the data.
The typical difference in length of $S$
for the relativistic and non-relativistic kinematics is less than
1-2~MeV, depending on the azimuthal opening angle,
$\phi_{12}$. This difference is less than the experimental resolution
in $S$ of 4~MeV (FWHM), and we, therefore, did not transform
the non-relativistic $S$-curves to the relativistic ones.

For the configurations at large azimuthal opening angle, $\phi_{12}\geq 40^\circ$,
and taking the systematic uncertainties into account, a reasonable agreement is observed between the cross-section data and the corresponding theoretical predictions. For the configuration with a small relative azimuthal angle, the picture changes.
Here, the measured cro\-ss sections show a large discrepancy with a calculation which includes the TM' 3NP. In this region, the CDB+$\Delta$+C\-o\-ulomb calculation has a smaller deficiency wh\-en compared with the experimental data, but the deficiency is still large for small value of polar angle, $\phi_{12}= 20^{\circ}$, as shown in Fig.~\ref{Eslami_fig_6}.

For the analyzing powers, the major discrepancies between the data and the
theoretical calculations arise at small azimuthal opening angles. In
this range, the predictions based solely on a NN potential are closest
to the data, although, the disagreement is still significant. The inclusion of
3NPs increases the gap between data and predictions as can be seen from bottom panels of Fig.~\ref{Eslami_fig_6}.
The contribution of the TM' 3NP appears to be larger than the implicit
inclusion of the $\Delta$ resonance by the Hannover-Lisbon theory group.
It is interesting to note that a similar,
but even larger, discrepancy has been observed in a break-up study at an
incident beam energy of 190~MeV~\cite{Hossein}.

This paper discusses some of the preliminary results of a proton-deuteron break-up experiment
carried out with an incident proton beam of 135~MeV. The data were taken using a nearly 4$\pi$ detection system,
BINA, and exploiting a polarized beam of protons. With this, precision differential cross sections
and analyzing powers were measured and compared to Faddeev calculations based on modern two- and
three-nucleon potentials. The large coverage of the detection system provides an ideal tool to
systematically explore the rich phase space of the break-up reaction. We have identified various
configurations at which significant discrepancies are observed between our data and predictions by
Faddeev calculations based upon state-of-the-art potentials. Intriguing deficiencies are observed for the analyzing power for
configurations at which the relative energy between the two outgoing protons becomes small.
The discrepancies cannot be explained by the Coulomb interaction
and higher-order relativity, since these effects are accounted for in the present state-of-the-art
calculations. Therefore, the data provide an ideal basis to develop a better understanding of
three-nucleon force effects in few-nucleon interactions.

\end{document}